† †

# Advection-Dominated Accretion Flows: Optically Thin Solutions

Ramesh Narayan[1,2]
1. Harvard-Smithsonian Center for Astrophysics, Cambridge, MA 02138
2. Institute for Theoretical Physics, University of California
Santa Barbara, CA 93106

## Abstract

General properties of advection-dominated accretion flows are discussed. Special emphasis is given to the optically thin branch of solutions, which has very high ion and electron temperatures and is thermally stable. This solution branch has been applied to a number of low-luminosity accreting black holes. The models have resolved some puzzles and have provided a straightforward explanation of the observed spectra. The success of the models confirms that the central objects in these low-luminosity sources are black holes. There is some indication that advection-dominated models may be relevant also for higher luminosity systems. The properties of the Low state of accreting black holes, the transition from the Low state to the High state, and the similarity of hard X-ray/$\gamma$-ray spectra of black hole X-ray binaries and active galactic nuclei, are explained.

## 1. Introduction

An advection-dominated accretion flow (ADAF) is defined as one in which a large fraction of the viscously generated heat is advected with the accreting gas, and only a small fraction of the energy is radiated. Although the basic idea of advection goes back a number of years (see the accompanying review by Abramowicz for a history of the subject), the relevance of ADAFs to real astrophysical systems was recognized only recently and much of the work in the subject dates back less than two years.

Advection-dominated accretion can occur in two different limits:

1. At very high mass accretion rates, radiation is trapped in the accreting gas because of the large optical depth and is advected with the flow. This limit of advection-domination, which typically occurs for mass accretion rates $\dot{M} > \dot{M}_{\rm Edd}$ (the Eddington rate), was considered initially by Begelman (1978) and Begelman & Meyer (1982). Abramowicz et al. (1988) carried out a detailed study of the solutions and discussed their properties, including stability.

2. At sufficiently low $\dot{M}$, the accreting gas can become optically thin. The cooling time of the gas is then longer than the accretion time, and once again we have an ADAF. This regime was first mentioned by Rees et al. (1982) in the context of their "ion torus" model, and has been studied in detail in recent papers by Narayan & Popham (1993), Narayan & Yi (1994, 1995a, b), Abramowicz et al. (1995), Chen (1995), and Chen et al. (1995).

In §2 of this review, we discuss the current status of our understanding of ADAFs in general. §3 then proceeds to a discussion of the particular features of the optically thin branch of solutions. §4 reviews applications of optically thin ADAFs to various astrophysical systems, and §5 concludes with some final remarks.

## 2. Structure of ADAFs

Ideally, we would like to understand the structure and properties of an ADAF in four dimensions, namely three coordinates ($R\theta\phi$) and time ($t$). Numerical simulations are the only way to achieve this goal, and given the enormous dynamic range involved and the messy radiation processes that one should include for a realistic calculation, no work has been done on the full 4D problem.

If we assume that the accretion flow is axisymmetric and in steady state, then we have a 2D problem ($R\theta$). Even this is a hard problem, because it leads to partial differential equations with boundary conditions. No solutions are available yet.

The work done so far on steady state ADAFs is limited to 0D and 1D models, with some attempt to approximate the 2D solution via a $(1+1)$D approach. The following subsections summarize the results. In addition, there have been preliminary studies of time-dependent flows in 2D ($Rt$, Manmoto, this volume) and 3D ($R\theta t$, Igumenshchev, Abramowicz & Chen 1995).

ADAF studies are currently at a stage roughly similar to that achieved in thin disk theory over the last 20 years. The bulk of the work on thin disks is based on the Shakura-Sunyaev (1973) model or its variants (cf. Frank, King & Raine 1992), which is a 0D solution in the language adopted here. Some attempts have been made to do 1D calculations to obtain the local vertical structure of thin disks (e.g. Hubeny 1990) or the global radial structure (e.g. Narayan & Popham 1993). In addition, there have been attempts to do time-dependent 2D ($Rt$) calculations of disk instabilities and 3D ($R\theta t$) computations of boundary layers in thin disks.

*2.1. 0D — The Basic Self-Similar Solution of ADAFs*

Narayan & Yi (1994) simplified the 2D axisymmetric steady state ADAF problem by assuming that the dynamical variables of the flow have power law dependences on $R$. This allowed them to evaluate directly all radial derivatives in the equations. Further, they eliminated $\theta$ by considering a height-integrated set of equations (the Slim Disk Equations, cf. Abramowicz et al. 1988). This results in a set of algebraic equations. The equations have an analytic self-similar solution which depends only on three parameters: the viscosity parameter $\alpha$, the ratio of specific heats of the accreting gas $\gamma$, and an advection parameter $f$ which is defined to be the ratio of the advected energy to the viscously generated energy. (Equivalently, the radiative efficiency is $1 - f$.) A variant of the Narayan & Yi solution had been derived earlier by Spruit et al. (1987) in a different context.

The self-similar solution reveals many of the basic properties of ADAFs:
(i) The radial velocity is $v \sim \alpha v_{\rm ff}$, where $v_{\rm ff} = (GM/R)^{1/2}$ is the free-fall velocity, with $M$ being the mass of the central object and $R$ the radius. The radial velocity is much larger than in thin disks.
(ii) The angular velocity is distinctly sub-Keplerian, $\Omega \lesssim \Omega_K/2$, the exact value depending on $\gamma$. Here $\Omega_K = (GM/R^3)^{1/2}$ is the Keplerian $\Omega$.
(iii) The isothermal sound speed is very large, $c_s \gtrsim v_{\rm ff}/2$.
(iv) The Bernoulli parameter of the accreting gas, viz. the sum of the kinetic energy, the potential energy and the enthalpy, is positive. Since the Bernoulli parameter is conserved in an adiabatic flow, if the gas in an ADAF moves adiabatically to a large radius it will have a bulk velocity comparable to the escape velocity at its point of origin. This suggests a possible connection between ADAFs and jets (Narayan & Yi 1995a) which needs to be explored.
(v) Entropy increases inward and, therefore, the accreting gas is convectively unstable. This has been confirmed by Igumenshchev et al. (1995) using numerical experiments (see Chen, this volume). It has been suggested that convection may possibly enhance the effective viscosity in ADAFs (Narayan & Yi 1994, 1995a).

Point (iii) implies that the temperature of the gas is nearly virial so that the vertical height of the gas is very large: $H \sim c_s/\Omega_K \sim R$. This leads one to ask the following important question: Is it valid to use height-integrated equations when the flow has such a thick morphology?

*2.2. 1D Solution in $\theta$ — Vertical Structure of ADAFs*

In a follow-up study, Narayan & Yi (1995a) avoided height-integration and considered an exact set of equations for a viscous axisymmetric steady state accretion flow. Assuming self-similarity in $R$, as had Begelman & Meyer (1982) earlier, they calculated exact numerical solutions of the resulting differential equations in $\theta$.

The solutions reveal that an ADAF is not at all disk-like in morphology, but nearly spherical. In fact, the flow configuration is similar to a differentially-rotating settling star, and there are no funnels. The radial velocity is maximum at the equator and goes to zero at the rotation pole, while the angular velocity and sound speed are nearly constant on spherical shells.

Despite the quasi-spherical morphology, the height-integrated analytic 0D solutions (§2.1) are in excellent agreement with the exact 1D solutions. This implies that the height-integration approximation is quite accurate even for nearly spherical flows. How is this possible? The answer is that height-integration should not be interpreted as a cylindrical average over $z$ at a given cylindrical radius $R$, but rather as a spherical average over $\theta$ at a given spherical radius $R$. With this new interpretation, the Slim Disk Equations are a valid description of the radial variations of both slim, cooling-dominated flows and quasi-spherical, ADAFs. This result provides the technical underpinning for much of the work on ADAFs, nearly all of which is based on height-integrated equations.

It should be emphasized that although an ADAF has a quasi-spherical morphology, it is nevertheless very different from pure spherical accretion (Bondi 1952). The ADAF solutions discussed here describe rotating flows where the accretion is completely controlled by angular momentum transport via viscosity. Bondi spherical accretion, on the other hand, ignores rotation altogether and the flow involves merely a competition between gravity and pressure. Since in nearly all applications we expect the accreting gas to have significant angular momentum, the ADAF solutions are more realistic than the classical spherical solutions.

*2.3. 1D Solution in R — Global Radial Solutions of ADAFs*

Recently, two groups have independently studied the global radial structure of ADAFs around black holes (Narayan, Kato & Honma 1996, Chen, Abramowicz & Lasota 1996). Using height-integrated equations (which are valid, see §2.2) they have obtained numerical solutions over a range of radius $R$ with consistent boundary conditions at both ends. On the inside, the gas flows through a sonic point and falls supersonically into the black hole. Early work on a similar problem was reported by Matsumoto, Kato & Fukue (1985).

These studies show that physically self-consistent, transonic, ADAF solutions are available for all values of $\alpha$ in the range $0 < \alpha \lesssim 0.3$. Therefore, there is no need for radial shocks in ADAFs. In contrast, transonic solutions do not appear to exist in thin accretion disks for $\alpha \gtrsim 0.05$, suggesting that thin disks may have shocks for such values of $\alpha$ (Chakrabarti 1990).

Another interesting result is that the 0D self-similar solution of Narayan & Yi (1994) is found to be quite a good representation of the global flow away from the boundaries. Near the sonic point, however, the radial velocity increases above the self-similar value, as does the angular velocity.

Global ADAF solutions with low values of $\alpha$ are found to be quite different from those with large $\alpha$. The former have sonic radii $R_s$ close to the marginally bound orbit, $R_s \sim 2R_g$ where $R_g = 2GM/c^2$ is the gravitational radius, while the latter have sonic radii close to or even outside the marginally stable orbit, $R_s \gtrsim 3R_g$.

Hydrostatic rotating thick tori with $R_s \sim 2R_g$ have been studied for many years as models of active galactic nuclei (Fishbone & Moncrief 1976, Abramowicz, Jaroszyński & Sikora 1978, Kozlowski, Jaroszyński & Abramowicz 1978, Paczyński & Wiita 1980, see Frank et al. 1992 for a simple discussion of the physics of these models). The models have (i) a region of super-Keplerian rotation near the inner edge, (ii) a radial pressure maximum within the disk which is the origin of the toroidal morphology, and (iii) twin empty funnels along the rotation axis. The funnels have been considered promising for the initiation of relativistic jets.

Global ADAF solutions with low values of $\alpha \lesssim 0.01$ are similar to thick tori in many respects. Because of the low viscosity, the radial velocity is small and there is a near hydrostatic balance between gravity, pressure and rotation. These flows have super-Keplerian rotation over a range of R and radial pressure maxima. Also, they have nearly empty funnels extending some distance along the rotation axis, though at large $R$ the funnels disappear and the flow becomes quasi-spherical as described in §2.2. These models are thus improved and self-consistent versions of the earlier tori, where now angular momentum and energy are conserved at each $R$.

Global ADAF solutions with large $\alpha$, however, appear to be completely different. They have sub-Keplerian rotation at all radii, have no pressure maxima, and show no sign of an axial funnel. These flows represent truly dynamical structures where, in contrast to low-$\alpha$ flows, the radial velocity is substantial. Viscosity strongly influences the dynamics when $\alpha$ is large, as Matsumoto et al. (1984) recognized even in the case of thin accretion disks.

Narayan, Kato & Honma (1996) extended the global height-integrated 1D solutions to approximate 2D solutions in $R\theta$ by means of a $(1+1)$D approach, where they used the local $\theta$ structure described in §2.2 to approximate the "vertical" structure at each $R$. The resulting isodensity contours confirm the quasi-spherical morphology and absence of funnels in large-$\alpha$ ADAFs.

*2.4. In Defence of the $\alpha$ Viscosity Prescription*

All the work on ADAFs so far has been based on the Shakura-Sunyaev $\alpha$ viscosity prescription. In this approach, the kinematic viscosity coefficient is written as $\nu = \alpha c_s^2/\Omega_K$ with $\alpha$ taken to be independent of $R$. A simpler approach is to set the shear stress equal to $\alpha P$, where $P$ is the pressure. In the opinion of this reviewer, the $\alpha$ prescription is well-motivated for an ADAF.

ADAFs have at least two linear instabilities, namely the Balbus-Hawley MHD instability (see articles by Balbus, Hawley and Achterburg in this volume) and a convective instability (Narayan & Yi 1994, 1995a, Igumenshchev et al. 1995). One of these instabilities (or perhaps both) doubtless produces the shear stress which causes angular momentum transport. The effective $\nu$ must, therefore, depend on the nonlinear saturation of the chosen instability. Now, an ADAF has only one length scale, namely the radius $R$; in contrast to thin disks, where $H$ is a second scale, here we have $H \sim R$. Further, there is only one velocity scale in the problem, namely $v_{\rm ff}$; the bulk flow of the gas has a speed $\sim v_{\rm ff}$, the sound speed is $c_s \sim v_{\rm ff}$, and if we have equipartition magnetic fields as seems reasonable with the Balbus-Hawley instability, then the Alfven speed too is $v_A \sim v_{\rm ff}$. This collapse of scales leads to an enormous simplification of the physics. Purely from dimensional analysis, it seems obvious that all linear instabilities must necessarily saturate with velocities $\sim v_{\rm ff}$ and coherence scales $\sim R$, so that the effective viscosity coefficient must necessarily scale as $\nu \propto v_{\rm ff} R \sim c_s^2/\Omega_K$. Further, the self-similar nature of the flow guarantees that the coefficient $\alpha$ in this relation will be independent of $R$. Therefore, the $\alpha$ prescription is particularly appropriate for ADAFs.

## 3. Optically Thin ADAFs

The rest of the article is devoted specifically to the optically thin branch of solutions. This section deals with general properties of the solutions, and the next section describes applications to a number of astrophysical objects.

*3.1 The Critical Mass Accretion Rate*

The optically thin branch is advection-dominated because of the poor radiative efficiency of the accreting gas. This requires the gas density to be sufficiently low that ion-electron coupling via Coulomb collisions (in a two-temperature plasma) becomes weak and cooling via synchrotron and bremsstrahlung radiation is unimportant. In addition, the optical depth has to be sufficiently low for Compton cooling to be modest. For these reasons, the optically thin branch exists only at low mass accretion rates.

In the following we scale radii in gravitational units, $r = R/R_g$, masses in solar units, $m = M/M_\odot$, and accretion rates in Eddington units, $\dot{m} = \dot{M}/\dot{M}_{\rm Edd} = \dot{M}/1.39 \times 10^{18} m$ g s$^{-1}$, where we have assumed a standard efficiency factor of 10% in defining the Eddington rate.

Various investigators (Abramowicz et al. 1995, Narayan & Yi 1995b, Chen et al. 1995) have estimated, under different assumptions, the maximum or critical accretion rate $\dot{m}_{\rm crit}$ up to which optically thin ADAFs exist. The value of $\dot{m}_{\rm crit}$ depends moderately on the radius ($\dot{m}_{\rm crit} \propto r^{-1/2}$ for large $r$), and quite strongly on the viscosity parameter ($\dot{m}_{\rm crit} \propto \alpha^2$). However, $\dot{m}_{\rm crit}$ is independent of $m$, and is insensitive to the magnetic field strength.

The most detailed calculations of $\dot{m}_{\rm crit}$ are due to Narayan & Yi (1995b). These calculations are based on a two-temperature plasma (Shapiro, Lightman & Eardley 1976), where all the energy from viscous dissipation first goes into the ions and then part of the energy is transferred to the electrons via Coulomb collisions. Cooling by bremsstrahlung and synchrotron radiation is included, along with Comptonization of these emissions. Pair processes are not included, but it has been checked a posteriori that the pair density is low (see Kusunose, this volume). All the gas parameters, including the ion temperature $T_i$ and the electron temperature $T_e$, are calculated self-consistently at each radius using the self-similar solution (§2.1), and the advection parameter $f$ is calculated at each radius self-consistently via the local cooling.

These calculations show that $\dot{m}_{\rm crit} \sim 0.3\alpha^2$ for $r \lesssim 10^3$ and $\dot{m}_{\rm crit} \sim 0.3\alpha^2(r/10^3)^{-1/2}$ for $r \gtrsim 10^3$. This means that unless $\alpha$ is fairly large, say $\alpha > 0.1$, optically thin ADAFs are restricted to extremely low $\dot{m}$ (see §4.6).

## 3.2 Properties of Optically Thin ADAFs

1. The accreting gas is extremely hot. The ion temperature is given by $T_i \sim 10^{12}{\rm K}/r$, which is nearly virial. The electron temperature at radii $r \lesssim 10^3$ is $T_e \sim 10^9 - 10^{10}$ K in the case of accreting black holes and $T_e \sim 10^{8.5} - 10^9$ K for accreting neutron stars (Narayan & Yi 1995b).

2. The accreting gas is thermally and viscously stable to long wavelength perturbations (Abramowicz et al. 1995, Narayan & Yi 1995b) and is only weakly unstable even to short wavelength perturbations (Kato, Abramowicz & Chen 1996, see Chen and Manmoto, this volume). Global stability is, therefore, assured.

3. The high electron temperature leads to a hard non-blackbody spectrum extending up to a few hundred keV. This, coupled with the thermal stability of the gas, makes these solutions extremely attractive for modeling accreting black holes (§4), most of which are known to have hard spectra in X-rays and soft $\gamma$-rays (Tanaka & Lewin 1995, Maisack et al. 1993). Previously, the only hot accretion solution known was the two-temperature cooling-dominated solution of Shapiro et al. (1976), which is thermally very unstable (Piran 1978) and is, therefore, unsuitable for realistic models.

4. Because these flows are advection-dominated, they are by definition inefficient radiators. Therefore, the flows are underluminous for their accretion rate. Whereas a cooling-dominated accretion flow around a black hole has a luminosity $L \sim \dot{m}L_{\rm Edd}$, optically thin ADAFs have $L \sim (\dot{m}^2/\dot{m}_{\rm crit})L_{\rm Edd}$. Thus, especially for $\dot{m} \ll \dot{m}_{\rm crit}$, the radiative efficiency is extremely low.

## 4. Astrophysical Applications of Optically Thin ADAFs

Because (i) optically thin ADAFs exist only for low values of $\dot{m}$, and (ii) the radiative efficiencies of these flows are low by construction, the most obvious applications are to low-luminosity systems. The majority of successful applications belong to this category, as we discuss in §4.3, 4.4 (see also articles by Lasota and Yi, this volume). Recently, there has been an attempt to explain luminous sources as well using these solutions and this is discussed in §4.5. So far, all applications have been limited to accreting black holes, but in principle the solutions could apply equally well to neutron stars (Narayan & Yi 1995b), and even to white dwarfs (for example, the hot inner flow in the siphon model of Meyer & Meyer-Hofmeister 1994 is advection-dominated).

*4.1 Spectral States of Accreting Black Holes*

It is well-known that black hole X-ray binaries (XRB) exhibit several distinct spectral states (Tanaka & Lewin 1995), and it is possible that active galactic nuclei (AGN) too have similar states. Before discussing specific applications, it is useful to describe a plausible scenario (Narayan 1995, 1996) in which we identify the role of advection-dominated flows in the various spectral states.

At the lowest luminosities, say $L < 10^{-4} L_{\rm Edd}$, we have the so-called Quiescent state or Off state. Soft X-ray transients (SXTs) exhibit this state in between outbursts, and very low luminosity AGN probably also correspond to this state. As we show in §4.3, 4.4, there is very good evidence that all systems in the Quiescent state have optically thin ADAFs, at least close to the black hole. The nature of the flow far from the black hole depends on the outer boundary condition. If the gas is introduced in a fairly hot or quasi-spherical state, then the flow would be advection-dominated at all $r$, as seems to be the case in low-luminosity AGN (e.g. Sagittarius A*, §4.3). However, if the incoming gas is cold and has a lot of angular momentum, then the flow would initially form a thin disk at large $R$. This is the case in SXTs. The observational evidence (§4.4) suggests that the accreting gas switches from a thin disk to an ADAF at some large transition radius $r = r_{\rm tr} \gg 1$. Two physical mechanisms have been suggested for the transition: the coronal siphon mechanism of Meyer & Meyer-Hofmeister (1994), and evaporation via diffusive heating as described by Honma (1995) (see articles by both authors in this volume).

With increasing $\dot{m}$, the flow configuration is likely to remain essentially the same as in the Quiescent state until $\dot{m}$ crosses $\dot{m}_{\rm crit}$. However, the luminosity will increase quite dramatically ($L \propto \dot{m}^2$, §3.2) and the system will become significantly brighter than in the Quiescent state. It has been suggested that systems with values of $\dot{m}$ approaching $\dot{m}_{\rm crit}$ correspond to the so-called Low State of accreting black holes (Narayan 1996). Observations indicate that sources in the Low state have luminosities $\sim (10^{-2} - 10^{-1}) L_{\rm Edd}$, very hard spectra with photon indices $\sim 1.5 - 2$, and no hint of any soft component in the spectrum that might correspond to a thin accretion disk. These properties are consistent with optically thin ADAFs with $\dot{m} \lesssim \dot{m}_{\rm crit}$ and $\alpha \sim 1$ (Narayan 1996, §4.5).

Once $\dot{m}$ crosses $\dot{m}_{\rm crit}$, optically thin ADAFs are no longer allowed, and the accreting gas has to switch to a different flow configuration; presumably, it becomes a pure thin accretion disk. Such a system would be quite luminous ($L \sim \dot{m} L_{\rm Edd}$), but with a soft multi-color blackbody spectrum. The High state of accreting black holes is known to match these properties. The change from the Low State to the High State happens quite suddenly according to the observations, and this is consistent with the ADAF model, where the transformation occurs when $\dot{m}$ crosses $\dot{m}_{\rm crit}$. The details of the transformation are probably as follows. When $\dot{m}$ approaches $\dot{m}_{\rm crit}$, we expect that the transition radius $r_{\rm tr}$ shrinks rapidly so that the outer thin disk progressively encroaches into the inner advection-dominated zone (Narayan 1996). During this stage, the spectrum would continue to be hard (though with a steepening power law index), and at the same time a soft component due to the thin disk would begin to increase in importance. Presumably, once $\dot{m}$ crosses $\dot{m}_{\rm crit}$, the thin disk extends all the way down to the black hole, and we have a full-fledged High state with a pure soft spectrum.

Finally, black hole systems also exhibit a so-called Very High state where the spectrum has both a luminous soft component and a substantial hard tail. It is claimed that this state corresponds to a higher $\dot{m}$ than the High state. The precise nature of the Very High state is not clear at this time. Perhaps, the flow consists of a standard thin disk and a corona, with the latter having a fairly substantial luminosity. The corona could be an ADAF in its own right (Narayan & Yi 1995b) or it may be efficiently cooled by Comptonization of disk photons (Haardt & Maraschi 1991).

*4.2 Modeling Procedures*

The models of individual objects published in the literature involve a few approximations. The most serious simplification is that the flow is assumed to have a locally self-similar form (§2.1), though the advection parameter $f$ is allowed to vary self-consistently with $r$ according to the local radiative efficiency (Narayan, McClintock & Yi 1996, Narayan 1996). Very recently, fully self-consistent models using the global flow described in §2.3 have been constructed (Narayan, Kato & Honma 1996, Chen, Abramowicz & Lasota 1996, Nakamura, this volume) and it has been confirmed that the self-similar approximation is quite good for calculating the spectrum.

In addition to the black hole mass $m$ and mass accretion rate $\dot{m}$, each model requires a choice of the viscosity parameter $\alpha$, the strength of the magnetic field (e.g. the parameter $\beta$ in Narayan & Yi 1995b), and the transition radius $r_{\rm tr}$ where the outer thin disk (if there is one) transforms to the inner advection-dominated flow. The plasma is assumed to be two-temperature (Shapiro et al. 1976), and thermal balance is enforced at each radius independently for the ions (viscous heating of ions balanced by Coulomb energy transfer to electrons) and the electrons (Coulomb heating balanced by radiative cooling). The radiative cooling is calculated with a radial Boltzmann-like radiative transfer code (rather than a full Monte Carlo calculation), with small corrections for non-spherical geometry and finite optical depth. The radiative transfer involves propagating the spectrum from one radial shell to the next, including emission via synchrotron and bremsstrahlung processes, Comptonization, and gravitational redshift. Electron positron pairs are included approximately, but their contribution is small.

The spectral calculations show that it is important to include synchrotron emission in the model since the Comptonization of synchrotron photons provides the bulk of the power in the X-ray band.

In the advection-dominated zone, the luminosity and spectrum depend only on two parameters, viz. $m$ and $\dot{m}/\alpha$, while in the outer thin disk, the relevant parameters are $m$, $\dot{m}$ and $r_{\rm tr}$. The results are insensitive to the assumed strength of the magnetic field (provided it is not vanishingly small).

*4.3 Sagittarius $A^*$ and Other Quiet AGN*

Sagittarius $A^*$ (Sgr $A^*$) at the center of the Galaxy is an unusual source which has remained a longstanding mystery. Dynamical evidence suggests that it has a mass $\sim 10^6 M_\odot$ which, combined with its unusual spectrum, suggests that the object is a supermassive black hole. Observations of gas flows in the vicinity of Sgr $A^*$ indicate a mass accretion rate onto the central object $\sim 10^{-4} M_\odot {\rm yr}^{-1}$ (Melia 1992, Genzel, Hollenbach & Townes 1994). In a standard disk with an efficiency of $\sim 10\%$, this accretion rate would correspond to a luminosity of $\sim 10^{42}$ ergs s$^{-1}$. In contrast, the actual luminosity observed is $\sim 10^{37}$ ergs s$^{-1}$. Further, the spectrum is essentially flat in $\nu L_\nu$ from radio to X-rays, with a few bumps. This is very different from the spectrum expected with a standard thin disk. Both the dimness of SgR $A^*$ and its peculiar spectrum were a puzzle until recently.

Narayan, Yi & Mahadevan (1995) showed that an optically thin ADAF model with a $7 \times 10^5 M_\odot$ black hole accreting at $\dot{M}/\alpha \sim 10^{-5} M_\odot {\rm yr}^{-1}$ fits the spectrum of Sgr $A^*$ (including all upper limits) quite well (see also Rees 1982). The predicted spectrum extends from radio to hard X-rays and even fits the observed spectral bumps reasonably well. Although $\dot{M}$ in the model is not quite as large as the direct estimate of $10^{-4} M_\odot {\rm yr}^{-1}$ mentioned above, it is nevertheless quite an improvement over the $\dot{M} \sim 10^{-9} M_\odot {\rm yr}^{-1}$ which one obtains with a standard thin disk.

The key feature of this model of Sgr $A^*$ is that the flow is advection-dominated, which explains the low luminosity of the source despite a large $\dot{M}$. Fabian & Rees (1995) have extended the idea to a number of nearby elliptical galaxies which are believed to have large nuclear black holes ($M \sim 10^8 - 10^9 M_\odot$). The black holes in these galaxies must accrete at reasonable mass accretion rates (Fabian & Canizares 1988) and yet they are unusually dim. Once again, an ADAF model provides a natural explanation.

Recently, Lasota et al. (1996, see Lasota, this volume) have developed a model of the nucleus of NGC 4258 involving an optically thin ADAF.

*4.4 Quiescent Soft X-ray Transients*

The Quiescent state of SXTs, in between outbursts, has been very difficult to understand in models involving thin accretion disks. The spectrum of the prototypical SXT, A0620-00, in quiescence consists of two components (McClintock, Horne & Remillard 1995): (i) an optical/UV component which is nearly blackbody in shape, and (ii) a weak X-ray tail. The optical/UV component is consistent with a thin accretion disk except that the inner edge of the disk has to be at $r > 10^3$ instead of at $r = 3$ as one expects for a standard thin disk. This is very unusual. Further, the X-rays cannot be produced by any standard thin disk model; if one attempts to fit the observed temperature then the luminosity becomes too large, while if one fits the X-ray luminosity then the temperature is too low. In addition, if one assumes that the X-rays are produced close to the black hole, then it is hard to understand why the X-ray luminosity is much less than the optical luminosity when the viscous dissipation is expected to have just the opposite behavior.

Narayan, McClintock & Yi (1996) developed a model for quiescent SXTs which resolves these puzzles and explains the observations. The flow consists of a thin disk on the outside which extends down to $r = r_{tr} \sim$ few $\times 10^3$, and an optically thin ADAF on the inside from $r = r_{tr}$ down to the black hole horizon at $r = 1$. The outer disk produces the optical and UV, while the ADAF produces the X-ray emission. The low luminosity of the X-rays is of course the result of advection-domination — nearly all the energy liberated by viscosity in the ADAF is advected into the black hole. Narayan et al. showed that they could fit the observed quiescent spectra of A0620-00, V404 Cyg and Nova Muscae 1991 quite well. Mineshige (1995) and Lasota, Narayan & Yi (1996) have shown that the thermal disk instability model of SXT outbursts (Mineshige & Wheeler 1989) is in much better agreement with observations if the thin disk is truncated at a large $r_{tr}$. This provides independent support for an ADAF at small radii in these systems.

*4.5 Luminous Sources*

On the face of it, an advection-dominated model is not an obvious choice for modeling a luminous black hole system since the high luminosity probably implies fairly efficient radiation. The reason for attempting it is that many luminous systems have substantial fluxes in hard X-rays and $\gamma$-rays of up to $100 - 200$ keV (Narayan 1995), and one requires a hot model with electron temperatures $T_e > 10^9$ K. The optically thin ADAF solution is currently the only known, self-consistent, dynamical model which is both hot and thermally stable. (Very recently, Esin et al. 1996 have discovered a cooling-dominated, hot, stable, one-temperature solution; applications of this solution have not yet been considered.)

Narayan (1996) has shown that advection-dominated models provide quite a good description of luminous black hole XRB and AGN, provided $\alpha$ is chosen to have a large value, $\alpha \sim 1$. The models correspond to $\dot{m}$ close to $\dot{m}_{\rm crit}$, where the radiative efficiency is moderately large. Spectral calculations give a good fit to observations in the Low State and provide a natural explanation for the transition from the Low State to the High State. The models also are consistent with the spectral evolution observed in SXTs during outburst.

The models indicate that $T_e$ is essentially independent of the black hole mass (cf. Narayan & Yi 1995b). This means that the X-ray/$\gamma$-ray spectra of black hole XRB and AGN should be qualitatively similar when they are in an advection-dominated state. Observations do suggest a close similarity. In contrast, a thin disk model predicts $T_{\rm eff} \propto m^{-1/4}$.

### 4.6 Constraint on $\alpha$

The application to luminous black hole systems discussed above requires a large value of $\alpha \sim 1$. The reason is that $\dot{m}_{\rm crit}$ is proportional to $\alpha^2$ (§3.1). If $\alpha$ is not large enough, then $\dot{m}_{\rm crit}$ becomes very small, and this means that even a system with the maximum allowed accretion rate, $\dot{m} = \dot{m}_{\rm crit}$, is just not luminous enough to match observations.

Interestingly, even applications to low-luminosity systems seem to suggest large values of $\alpha$. In Sgr A*, the best agreement between the model $\dot{M}$ and the directly estimated $\dot{M}$ is obtained with $\alpha \sim 1$. Similarly, in quiescent SXTs, the models work best with $\alpha \sim 0.1 - 0.3$, and there are analogous indications in NGC 4258 as well.

All of these results have been obtained using the self-similar solution. It is possible that when the spectral calculations are done with a fully global model (§2.3) the actual value of $\alpha$ needed may not quite be unity but may be somewhat smaller. However, it is not likely to reduce below $\sim 0.1 - 0.3$.

Simulations of MHD instabilities in accretion flows generally do not give large values of $\alpha$ but rather give $\alpha \sim 0.01$ (see articles by Hawley and Achterburg, this volume). The simulations, however, correspond to standard thin disks. If $\alpha$ increases with increasing $H/R$, then it is conceivable that $\alpha \gtrsim 0.1$ may be allowed in ADAFs which have $H/R \sim 1$.

### Conclusion

To conclude, the optically thin ADAF solution provides for the first time a thermally stable and dynamically self-consistent flow model which has the extremely high electron temperatures ($T_e > 10^9$ K) needed to explain X-ray/$\gamma$-ray observations of accreting black holes (and some neutron stars). Spectra calculated with these models resemble observations quite well. The models work convincingly in low-$\dot{m}$ systems (§4.3, 4.4) and show promise even for high-$\dot{m}$ systems, provided $\alpha$ is large (§4.5).

The detailed models developed so far are based on a two-temperature plasma. One may interpret the success of the models as empirical support for the two-temperature idea (Narayan, McClintock & Yi 1996).

Perhaps the most exciting aspect of these models is that advection is not merely a minor perturbation but is in some sense the whole story. Especially for the applications discussed in §4.3, 4.4, it is only because of massive advection of viscous energy into the black hole that the models work at all. The existence of a horizon is critical since it ensures that whatever energy falls into the central object disappears without being re-radiated. It is not enough just to have a relativistic central star — we specifically need a *black hole*. The success of the models in the various applications discussed in this article could, therefore, be considered as "proof" that these particular systems do have central black holes.

## Acknowledgements

The author thanks T. Manmoto and I. Yi for comments on the manuscript. This work was supported in part by NASA grant NAG 5-2837 (to the Harvard-Smithsonian Center for Astrophysics) and NSF grant PHY 9407194 (to the Institute for Theoretical Physics).